\documentstyle[12pt]{article}
\begin{document}
\thispagestyle{empty}

\noindent
\hfill  OHSTPY-HEP-T-92-014      \\
\hfill hep-th/9210147   \\
\phantom{00} \hfill DOE/ER/01545-584\\
\vspace{1cm}
\begin{center}
  \begin{Large}
  \begin{bf}
    Renormalization of Tamm-Dancoff Integral Equations
   \\
  \end{bf}
 \end{Large}
\end{center}
  \vspace{.75cm}
    \begin{center}
\begin{large}
Brett van de Sande\\[8pt]
Stephen S. Pinsky\\
      \vspace{0.4cm}
\end{large}
\begin{it}
Department of Physics\\
The Ohio State University\\
174 West 18th Avenue\\
Columbus, Ohio  43210\\
\end{it}
  \end{center}
  \vspace{1cm}
\baselineskip=35pt
\begin{abstract}
\noindent
During the last few years, interest has arisen in using light-front
Tamm-Dancoff
field theory to describe relativistic bound states for theories such as QCD.
Unfortunately, difficult renormalization problems stand in  the way.  We
introduce a general, non-perturbative approach to renormalization that is well
suited for the ultraviolet and, presumably, the infrared  divergences found in
these systems.  We reexpress the renormalization problem in terms  of a set of
coupled inhomogeneous integral equations, the ``counterterm equation.''

The solution of this equation provides a kernel for the Tamm-Dancoff integral
equations which generates states that are independent of any cutoffs.  We also
introduce a Rayleigh-Ritz approach to numerical solution of  the
counterterm equation. Using our approach to renormalization, we examine several
ultraviolet divergent models.  Finally, we use the  Rayleigh-Ritz approach to
find the counterterms in terms of allowed operators of a theory.
\end{abstract}
\newpage

\begin{large}
\begin{bf}
\noindent
Introduction
\end{bf}
\end{large}
\vspace{.1in}
\baselineskip=18pt

First attempted in the 1950's by Tamm and Dancoff \cite{tamm,dancoff}, the
idea of describing relativistic bound states in terms of a small number of
particles (Fock space truncation) ran into considerable difficulties.  There
were  severe divergences associated with connected Green's functions as well as
divergences associated with particle creation and annihilation from the vacuum
and the approach was soon abandoned.  In 1966, Weinberg noticed that, in the
infinite momentum frame, creation and annihilation of particles from the vacuum
is forbidden and the divergences associated with the vacuum itself are removed
\cite{weinberg}.  This is similarly true for light front quantized field
theories.  During the last few years, interest has arisen in combining these
two
ideas, Tamm-Dancoff Fock space truncation and light front quantization,  in an
attempt to describe relativistic bound states \cite{perry,brodskylepage}.
Although divergences associated with the vacuum are removed, very difficult
renormalization problems remain.

In perturbation theory, divergences fall into two categories:  ultraviolet
(UV) divergences and infrared (IR) divergences.  One generally associates UV
divergences with particles having large energy.  In the context of light-front
physics, large energies arise from particles having large transverse or small
longitudinal momenta, and thus UV divergences and some IR divergences have a
common origin.  New non-perturbative divergences arise because the Tamm-Dancoff
truncation does not include all diagrams for any given order in perturbation
theory.  To date, these divergences  in light-front Tamm-Dancoff calculations
have only been renormalized within  the context of perturbation theory
\cite{yukawa}.  What we propose here is a general, nonperturbative approach to
renormalization that naturally handles the UV and, we believe,  some of the IR
divergences found in light-front Tamm-Dancoff field theory.

In this approach, it is shown that the renormalization problem can be rewritten
as a set of coupled integral equations, much in the same way that the
eigenvalue
problem is written as a set of coupled integral equations.  One can then solve
this set of coupled integral equations numerically using the same techniques
that are used in solving the eigenvalue problem itself.  Although the context
of
light-front field theory is the motivation for this work,  our approach will
not
depend on any of its special properties.  Presumably, our approach to
renormalization could be used in other, unrelated contexts.

In section 1, we will introduce the renormalization problem using a simple one
dimensional UV divergent model that has been discussed by other authors.  We
will introduce the so-called ``high-low analysis'' and show how it can be used
to renormalize the model.  In section 2 we will use the high-low
analysis to examine the general case and derive the ``counterterm equation,''
a set of coupled integral equations which relate the bare and renormalized
Hamiltonians.  In section 3, we will show how renormalization group ideas are
expressed in the context of our renormalization procedure. Next, in section 4,
we will discuss a variational, or Rayleigh-Ritz, approach  to solving the
counterterm equation.  In sections 5 and 6, we will apply our renormalization
scheme to some simple examples.  Finally, in section 7,  we use the
Rayleigh-Ritz approach to find the counterterm in terms of the allowed
operators
of a theory.\\

\begin{large}
\begin{bf}
\noindent
1.  Model A
\end{bf}
\end{large}

\vspace{.1in}
Let us start by looking at model A, a simple toy model that has been studied
by a number of authors \cite{yukawa,moriand,thorn,jackiw}.  Consider the
homogeneous integral equation

$$\left(p - E\right) \phi\left(p\right) + g \int_0^\Lambda
dp^\prime\;\phi\left(p^\prime\right) = 0\eqno (1.1)$$

\vspace{.1in}
\noindent
with eigenvalue $E$  and eigenvector $\phi\left(p\right)$.  This is a model
for a single particle of momentum $p$ with Hamiltonian
$H\left(p,p^\prime\right)
= p\;\delta\left(p - p^\prime\right) + g$.  We will focus on the $E < 0$
solution:

$$ \phi\left(p\right) = {{\rm const}\over {p - E}}\;,\;\;\;\;\;\; E =
{\Lambda\over{1 - e^{-1/g}}}\;. \eqno(1.2)$$

\vspace{.1in}
\noindent
Note that the eigenvalue diverges in the limit $\Lambda\to\infty$.  Proper
renormalization involves modifying the system to make $E$ and
$\phi\left(p\right)$ independent of $\Lambda$ in the limit $\Lambda\to\infty$.
Toward this end we add a counterterm $C_\Lambda$ to the Hamiltonian. Invoking
the high-low analysis \cite{wilson}, we divide the interval $0<p<\Lambda$ into
two subintervals:  $0<p<L$, a ``low-momentum region,''  and $L<p<\Lambda$, a
``high-momentum region,'' where the momentum scales characterized by $E$, $L$,
and $\Lambda$ are assumed to be widely separated. The idea is that the
eigenvalue and eigenvector should be independent of  the behavior of the system
in the high-momentum region.  The eigenvalue equation can be written as two
coupled equations

$$p\ \epsilon\left[0,L\right]\;\;\;\; \left(p - E\right) \phi\left(p\right)
+ \left(g + C_\Lambda\right) \int_0^L dp^\prime\;\phi\left(p^\prime\right) +
\left(g + C_\Lambda\right) \int_L^\Lambda dp^\prime\;\phi\left(p^\prime\right)
= 0\eqno (1.3a)$$

$$p\ \epsilon\left[L,\Lambda\right]\;\;\;\; \left(p - E\right)
\phi\left(p\right) + \left(g + C_\Lambda\right) \int_L^\Lambda dp^\prime\;
\phi\left(p^\prime\right) + \left(g + C_\Lambda\right) \int_0^L
dp^\prime\;\phi\left(p^\prime\right) = 0\;.\eqno (1.3b)$$

\vspace{.1in}
\noindent
Integrating (1.3b) in the limit $L,\Lambda\gg E$,

$$\int_L^\Lambda dp\;\phi\left(p\right) = -{{\left(g + C_\Lambda\right) \ln
{\Lambda\over L}}\over{1+ \left(g + C_\Lambda\right) \ln {\Lambda\over L}}}
\int_0^L dp^\prime\;\phi\left(p^\prime\right)\;,\eqno (1.4)$$

\vspace{.1in}
\noindent
and substituting this expression into (1.3a), we obtain an eigenvalue equation
with the high momentum region integrated out:

$$p\ \epsilon\left[0,L\right]\;\;\;\; \left(p - E\right) \phi\left(p\right)
+ {{\left(g + C_\Lambda\right)}\over{1+ \left(g + C_\Lambda\right) \ln
{\Lambda\over L}}}\int_0^L dp^\prime\;\phi\left(p^\prime\right) = 0\;.\eqno
(1.5)$$

\vspace{.1in}
\noindent
If we demand that this expression be independent of $\Lambda$,

$$ {d\over{d\Lambda}}\left({{\left(g + C_\Lambda\right)}\over{1+ \left(g +
C_\Lambda\right) \ln {\Lambda\over L}}}\right) = 0\;,\eqno (1.6)$$

\vspace{.1in}
\noindent
we obtain a differential equation for $C_\Lambda$

$${{d C_\Lambda}\over{d\Lambda}} = {{\left(g +
C_\Lambda\right)^2}\over\Lambda}\;.\eqno (1.7)$$

\vspace{.1in}
\noindent
Solving this equation, we are free to insert an arbitrary constant
$-1/A_\mu - \ln\mu$

$$g + C_\Lambda = {A_\mu\over{1 - A_\mu \ln{\Lambda\over\mu}}}\;.\eqno (1.
8)$$

\vspace{.1in}
\noindent
Substituting this result back into (1.5),

$$p\ \epsilon\left[0,L\right]\;\;\;\; \left(p - E\right) \phi\left(p\right) +
{{A_\mu}\over{1- A_\mu \ln {L\over \mu}}}\int_0^L
dp^\prime\;\phi\left(p^\prime\right) = 0\eqno (1.9)$$

\vspace{.1in}
\noindent
we see that $\Lambda$ has been removed from the equation entirely.  Using
(1.8) in the original eigenvalue equation

$$\left(p - E\right) \phi\left(p\right) + {A_\mu\over{1 - A_\mu
\ln{\Lambda\over\mu}}} \int_0^\Lambda dp^\prime\;\phi\left(p^\prime\right)
= 0 \eqno (1.10)$$

\vspace{.1in}
\noindent
gives the same equation as (1.9) with $L$ replaced by $\Lambda$.  The
eigenvalue is now,
$$E = {\Lambda\over{1 -{\Lambda\over \mu} e^{-1/A_\mu}}}\;\;\;\;\;
\lim_{\Lambda\to\infty}\; E = - \mu e^{1/A_\mu}\;.\eqno (1.11)$$

\vspace{.1in}
\noindent
Although the eigenvalue is still a function of the cutoff for finite $\Lambda$,
the eigenvalue does become independent of the cutoff in the limit
$\Lambda\to\infty$, and the system is properly renormalized.

We can think of $A_\mu$ as the renormalized coupling constant and $\mu$ as the
renormalization scale.  In that case, the eigenvalue should depend on the
choice
of $A_\mu$ for a given $\mu$ but be independent of $\mu$ itself.  Suppose, for
equation (1.11), we want to change $\mu$ to a new value, say $\mu^\prime$.  In
order that the eigenvalue remain the same, we must also change the coupling
constant from $A_\mu$ to $A_{\mu^\prime}$

$$\mu e^{1/A_\mu} = \mu^\prime e^{1/A_{\mu^\prime}}\;.\eqno (1.12)$$

\vspace{.1in}
\noindent
In the same manner, one can write down a $\beta$ function for $A_\mu$

$$\mu\;{d\over{d\mu}} A_\mu  = A_\mu^2\;.\eqno(1.13)$$\\

\begin{large}
\begin{bf}
\noindent
2.  The Counterterm Equation
\end{bf}
\end{large}

\vspace{.1in}
Using the ideas introduced in the previous section we can examine the general
case.  Throughout, we will be working with operators projected onto some
Tamm-Dancoff subspace (finite particle number) of the full Fock space.  In
addition, we will regulate the system by demanding that each component of
momentum of each particle lies within some finite interval.  We define the
``cutoff'' $\Lambda$ to be an operator which projects onto this subspace of
finite particle number and finite momenta.  Thus, for any operator $O$, $O
\equiv \Lambda O \Lambda$.  Let us introduce the Hamiltonian

$$H = H_0 + V + C_\Lambda\eqno (2.1)$$

\vspace{.1in}
\noindent
where, in the standard momentum space basis, $H_0$ is the diagonal part of
the Hamiltonian, $V$ is the interaction term, and $C_\Lambda$ is the
counterterm which is to be determined and is a function of the cutoff.  Each
term of the Hamiltonian is Hermitian and compact.  Schr\"{o}dinger's equation
can be written

$$\left(H_0 - E \right) \phi + \left( V + C_\Lambda\right)\phi = 0\eqno (2.2)$$

\vspace{.1in}
\noindent
with energy eigenvalue $E$ and eigenvector $\phi$.  Our goal is to choose
$C_\Lambda$ such that $E$ and $\phi$ are independent of $\Lambda$ in the
limit of large cutoff.

Now we will make an important assumption:  the physics that we are interested
in, characterized by energy scale $E$, is independent of physics near the
boundary of the space spanned by $\Lambda$.  Thus, we define two projection
operators, ${\cal H}$ and ${\cal L}$, where $\Lambda = {\cal H} + {\cal L}$,
${\cal H L} = {\cal L H} = 0$, and ${\cal H}$ and ${\cal L}$ commute with
$H_0$.    ${\cal H}$ projects onto a ``high-momentum region'' which contains
energy scales we do not care about, and ${\cal L}$ projects onto a
``low-momentum
region'' which contains energy scales characterized by $E$. ~Schr\"{o}dinger's
equation (2.2) can be rewritten as two coupled equations:

$$\left(H_0 - E \right){\cal L}\phi + {\cal L}\left( V + C_\Lambda \right)
{\cal L}\phi  + {\cal L}\left( V + C_\Lambda\right){\cal H}\phi = 0 \eqno
(2.3a)$$

\vspace{.1in}
\noindent
and

$$\left(H_0 - E \right){\cal H}\phi + {\cal H}\left( V + C_\Lambda \right){\cal
H}\phi  + {\cal H}\left( V + C_\Lambda\right){\cal L}\phi = 0\;. \eqno (2.3b)$$

\vspace{.2in}
\noindent
Using equation (2.3b), we can formally solve for $\cal H\phi$ in terms of $\cal
L\phi$

$${\cal H}\phi = {1\over {{\cal H}\left(E - H\right){\cal H}}} \left( V +
C_\Lambda \right){\cal L}\phi\;.\eqno (2.4)$$

\vspace{.1in}
\noindent
The term with the denominator is understood to be defined in terms of its
series expansion in $V$.  We can substitute this result back into (2.3a)

$$\left(H_0 - E \right){\cal L}\phi + {\cal L}\left( V + C_\Lambda \right){\cal
L}\phi  + {\cal L}\left( V + C_\Lambda\right) {1\over {{\cal H}\left(E -
H\right){\cal H}}} \left( V + C_\Lambda \right){\cal L}\phi = 0\;.\eqno (2.5)$$
\\

In order to properly renormalize the system, we could choose $C_\Lambda$ such
that (2.5) is independent of one's choice of $\Lambda$ for a fixed ${\cal L}$
in the limit of large cutoffs.  However, we will make a stronger demand:
that (2.5) should be equal to (2.2) with the cutoff $\Lambda$ replaced by
${\cal L}$.  Consider the following ansatz for $C_\Lambda$:

$$C_\Lambda = a_0 V + a_1 V F V + a_2 V F V F V + \ldots \eqno (2.6)$$

\vspace{.1in}
\noindent
where $F$ commutes with ${\cal H}$ and ${\cal L}$ and the coefficients $a_0,
a_1, \ldots\ $ are to be determined.  We will discuss some specific choices
for $F$ below.  Substituting (2.6) into (2.2) with $\Lambda$ replaced by
${\cal  L}$ everywhere and setting it equal to (2.5), order by order in $V$,

\begin{eqnarray*}
\lefteqn{\left(1 + a_0\right) V +  a_1  V{\cal L}F V + a_2  V{\cal L}F V{\cal
L} F V + \ldots}\\
& & = \left(1 + a_0\right) V + a_1 V F V + a_2 V F V F V\\
& & + \left(1 + a_0\right)^2 V{{\cal H}\over{E - H_0}}V + a_1 \left(1 +
a_0\right)\left(V F V{{\cal H}\over{E - H_0}} V + V{{\cal H}\over{E - H_0}}
V F V\right)\\
& & + \left(1 + a_0\right)^3 V{{\cal H}\over{E - H_0}}V{{\cal
H}\over{E - H_0}}V + \ldots\;.\end{eqnarray*}
\vspace{-.6in}
$$\eqno (2.7)$$
\\
\noindent
Using  $\Lambda = \cal L + \cal H$, we find that $a_0$ is arbitrary and the
rest of the coefficients are uniquely determined

$$a_i = (-1)^i (1 + a_0)^i\;,\;\;\;i = 1, 2, \ldots\eqno (2.8)$$

\vspace{.1in}
\noindent
provided that we can make the approximation

$$V {{\cal H}\over{E - H_O}}V \approx  V{\cal H} F V\;.\eqno (2.9)$$

\vspace{.1in}
\noindent
This is what we will call the ``renormalizability condition.''  A system is
properly renormalized if, as we increase the cutoffs $\Lambda$ and ${\cal L}$,
(2.9) becomes an increasingly good approximation.  Since the choice of $a_0$
rescales $V$, we absorb the factor of $1 + a_0$ into the definition of $V$. The
arbitrary scale $1 + a_0$ corresponds to a particular solution of (2.7) and
therefore depends on $F$, where $F$ may involve an arbitrary energy scale
$\mu$.
Thus $1 + a_0$ is an implicit function of $\mu$.  In simple models where $V$ is
proportional to a coupling constant, this $\mu$-dependent rescaling is
equivalent to introducing a running coupling $g(\mu)$ that runs with the
arbitrary scale in $F$.  In both models A and B discussed below the coupling
will therefore run with $\mu$.  Summing the series (2.6), we can express
$C_\Lambda$
as the solution of an  operator equation, the ``counterterm equation,''

$$C_\Lambda  = - V F V -  V F C_\Lambda \eqno (2.10)$$

\vspace{.1in}
\noindent
or, defining $V_\Lambda  = V + C_\Lambda$,

$$V_\Lambda = V -  V F V_\Lambda\;.\eqno (2.11)$$

\vspace{.1in}
\noindent
This is our central result.  In the standard momentum space basis, this
becomes a set of coupled inhomogeneous integral equations.  Such equations
generally have a unique solution, allowing us to renormalize systems without
having to resort to perturbation theory.  This includes cases where the
perturbative  expansion diverges or converges slowly.

There are many possible choices for $F$ that satisfy the renormalizability
condition.  For instance, one might argue that we want $F$ to resemble
$1/\left(E - H_0\right)$ as much as possible and choose

$$F = {1\over{\mu  - H_0}}\eqno (2.12)$$

\vspace{.1in}
\noindent
where the arbitrary constant $\mu$ is chosen to be reasonably close to $E$. In
this case, one might be able to use a smaller cutoff in numerical calculations.

One might argue that physics above some energy scale $\mu$ is simpler and that
it is numerically too difficult to include the complications of the physics at
energy scale $E$ in the solution of the counterterm equation.  Thus one could
choose

$$F = -{{\theta\left(H_0 - \mu\right)}\over{H_0}}\eqno (2.13)$$

\vspace{.1in}
 \noindent
where the arbitrary constant $\mu$ is chosen to  be somewhat larger than $E$
but smaller than the energy scale associated with  the cutoff.  The
$\theta$ function is assumed to act on each diagonal element in the standard
momentum space basis.  The difficulty with this renormalization scheme is that
it involves three different energy scales, $E$, $\mu$ and the cutoff which
might
make the numerical problem more difficult.\\

\begin{large}
\begin{bf}
\noindent
3.  Renormalization-Group Analysis
\end{bf}
\end{large}

\vspace{.1in}
We can relate our approach to conventional renormalization-group concepts.
In renormalization-group language, $V_\Lambda$ is the bare interaction term and
$V$ is the renormalized interaction term.  In both of the renormalization
schemes introduced above, we introduced an arbitrary energy scale $\mu$; this
is the renormalization scale.  Now, physics (the
energy eigenvalues and eigenvectors) should not depend on this parameter or
on the renormalization scheme itself, for that matter.  How does one move from
one renormalization scheme to another?  Consider a particular choice of
renormalized interaction term $V$ associated with a renormalization scheme
which uses $F$ in the counterterm equation.  We can use the counterterm
equation to find the bare coupling $V_\Lambda$ in terms of $V$.  Now, to find
the renormalized interaction term $V^\prime$ associated with a different
renormalization scheme using a different operator $F^\prime$ in the
counterterm equation, we simply use the counterterm equation with $V_\Lambda$
as given and solve for $V^\prime$

$$V^\prime = V_\Lambda  +  V_\Lambda F^\prime V^\prime\;.\eqno (3.1)$$

\vspace{.1in}
\noindent
Expanding this procedure order by order in $V$ and summing the result, we can
obtain an operator equation relating the two renormalized interaction terms
directly:

$$V^\prime  = V + V\left(F^\prime - F\right) V^\prime\;.\eqno (3.2)$$

\vspace{.1in}
\noindent
The renormalizability condition ensures that this expression will be
independent of the cutoff in the limit of large cutoff.

For the two particular renormalization schemes mentioned above, (2.12)
and (2.13), we can regard the renormalized interaction term $V$ as an implicit
function of $\mu$.  We can see how the renormalized interaction term
changes with $\mu$ in case (2.12):

$$\mu\;{d\over{d\mu}}V  = - V{{\mu}\over{\left(H_0 - \mu\right)^2}}V\eqno
(3.3)$$

\noindent
and in the case (2.13):

$$\mu\;{d\over{d\mu}}V  = V\delta\left(H_0 - \mu\right)V\;.\eqno (3.4)$$

\vspace{.1in}
\noindent
This is a generalization of the $\beta$ function.\\

\begin{large}
\begin{bf}
\noindent
4.  The Rayleigh-Ritz Method
\end{bf}
\end{large}

\vspace{.1in}
It is necessary to find efficient numerical techniques for solving the
counterterm equation.  The approach we discuss here is a simple extension
of a standard technique used in solving Fredholm-type integral equations
\cite{porterstirling}.  Suppose we have some operator $K$; how do we measure,
given $V$ and $F$, whether $K$ is a good approximation to $V_\Lambda$?  Let us
define a functional $J\left[ K\right]$ such that $J\left[ K\right]$ is
stationary at $K = V_\Lambda$.  The expression

$${\rm Tr}\; \left(K - V_\Lambda\right)\left(F + F V F\right)\left(K -
V_\Lambda\right) \eqno (4.1)$$

\vspace{.1in}
\noindent
is stationary at $K = V_\Lambda$.  Using the counterterm equation (2.10),
$\left(F + F V F\right)V_\Lambda = F V$, and (4.1) can be rewritten as

$${\rm Tr}\; K\left(F + F V F\right)K - {\rm Tr}\; K F V - {\rm Tr}\; V F K +
{\rm term\;independent\;of\;K}\;.\eqno (4.2)$$

\vspace{.1in}
\noindent
Thus, the functional

$$J\left[ K\right] = {\rm Tr}\; K\left(F + F V F\right)K - {\rm Tr}\; K
\left(F V + V F\right)\eqno (4.3)$$

\vspace{.1in}
\noindent
is stationary at $K = V_\Lambda$ \cite{general}.  Let us choose a linearly
independent set of Hermitian operators $\left\{O_i \right\}$ and define

$$K = \sum_i\;x_i\;O_i\;.\eqno(4.4)$$

\vspace{.1in}
\noindent
We want to find the vector ${\bf x}$ such that $J\left[ K\right]$ is
stationary; this will give us the best approximation to $V_\Lambda$ from the
subspace spanned by $\left\{O_i \right\}$.  Defining the Hermitian matrix $B$
and  the real-valued vector ${\bf c}$

$$B_{i,j} = {\rm Tr}\; O_i \left(F + F V F\right) O_j \;\;\;\;\;\; c_i = {\rm
Tr}\; O_i\left(F V +V F\right)\;,\eqno(4.5a,b)$$

\vspace{.1in}
\noindent
the functional is

$$J\left[ K\right] = {\bf x}^\dagger\;B\;{\bf x} - {\bf x}\cdot{\bf c}\eqno
(4.6)$$

\vspace{.1in}
\noindent
with stationary point given by

$$2 B\;{\bf x} = {\bf c}\;.\eqno (4.7)$$

\vspace{.1in}
\noindent
Unless the number of operators is limited, this approach is not very  useful.
If we can approximate the space spanned by $\Lambda$ using $n$ vectors, then
it will generally take $n^2$ operators to span the truncated Fock space;
finding
${\bf x}$ involves solving an $n^2 \times n^2$ linear system, too large for
practical applications.

Alternatively, we can express $K$ as an outer product of $n$ linearly
independent vectors $\left\{\chi_i\right\}$ \cite{mo}

$$K = \sum_{i,j = 1}^n\; X_{i,j}\;\chi_i\;\chi_j^\dagger\;.\eqno (4.8)$$

\vspace{.1in}
\noindent
Define the Hermitian matrices $A$, $E$, and $C$:

\begin{eqnarray*}
A_{i,j} &=& \chi_i^\dagger \left(F + F V F\right) \chi_j\\
E_{i,j} &=& \chi_i^\dagger\;\chi_j\\
C_{i,j} &=& \chi_i^\dagger\left(F V +V F\right)\chi_j
\end{eqnarray*}

\vspace{-.6in}
$$\eqno (4.9a,b,c)$$

\vspace{.1in}
\noindent
Note that $E$ is positive definite.  The functional is now written as

$$J\left[ K\right] = {\rm Tr}\; X E X A - {\rm Tr}\; C X\;;\eqno (4.10)$$

\vspace{.1in}
\noindent
its stationary point is given by the matrix equation

$$C = A X E + E X A\;.\eqno (4.11)$$

\vspace{.1in}
\noindent
One can solve for $X$ using standard numerical techniques; the solution is
basically equivalent to solving an $n \times n$ generalized eigenvalue
problem \cite{golubvanloan,solution}.\\

\begin{large}
\begin{bf}
\noindent
5.  Model A Revisited
\end{bf}
\end{large}

\vspace{.1in}
Model A is an example of what could be called a ``separable system.''
Separable systems have interaction terms which can be written as an outer
product of a Fock space vector:  $V = \eta \eta^\dagger$.  Then, the
counterterm equation is

$$V_\Lambda = \eta \eta^\dagger - \eta \eta^\dagger F V_\Lambda \eqno (5.1)$$

\vspace{.1in}
\noindent
which has the solution

$$V_\Lambda = {{\eta \eta^\dagger}\over{1 + \eta^\dagger F \eta}}\;.\eqno
(5.2)$$

\vspace{.1in}
\noindent
So, in this case, the counterterm equation produces an ordinary coupling
constant renormalization.  For model A in the standard momentum space basis,
we have $\eta\left(p\right) = {\sqrt g}$ where $g = g\left(\mu\right)$, and the
bare interaction term is

$$V_\Lambda\left(p,p^\prime\right) = {g\over{1 + g \int_0^\Lambda
dq\;dq^\prime\;  F\left(q,q^\prime\right)}}\;.\eqno (5.3)$$

\vspace{.1in}
\noindent
For the choice $F\left(q,q^\prime\right) = \delta\left(q -
q^\prime\right)/\left(\mu  - q\right)$ the eigenvalue is, using a
principal-value prescription,

$$E = {\Lambda\over{1 - {\textstyle{e^{-1/g}\left(\Lambda - \mu\right)}\over
\textstyle{\left|\mu\right|}}}}\eqno (5.4)$$

\vspace{.1in}
\noindent
which, in the limit $\Lambda\to\infty$ becomes independent of $\Lambda$,
giving an expression consistent with our previous result,

$$\lim_{\Lambda\to\infty} E = -\left|\mu\right| e^{1/g}\;.\eqno (5.5)$$

\vspace{.1in}
\noindent
We can easily verify the renormalizability condition for this model.  For the
${\cal H}$ region, we restrict $L<p<\Lambda$.  The renormalizability
condition is,

$$g \ln\left({{\Lambda-E}\over{L-E}}\right) \approx  g \ln
\left({{\Lambda-\mu}\over{L-\mu}}\right) \eqno (5.6)$$

\vspace{.1in}
\noindent
which becomes an equality in the limit $L,\Lambda \to\infty$.  Finally, we can
use equation (3.3) to calculate the $\beta$ function:

$$\mu\;{d\over{d\mu}}\ g\left(\mu\right)  = {g\left(\mu\right)}^2 \left(1 +
{\mu\over{\Lambda - \mu}}\right)\;\;\;\;\;\lim_{\Lambda\to\infty}
\mu\;{d\over{d\mu}}\ g\left(\mu\right) = {g\left(\mu\right)}^2\;.\eqno (5.7)$$

\vspace{.1in}
\noindent
which is the same result we found in (1.13).\\

\begin{large}
\begin{bf}
\noindent
6.  Model B
\end{bf}
\end{large}

\vspace{.1in}
Let us consider an extension to model A that has a more complicated
interaction term.  In the momentum space basis, we define the Hamiltonian of
model B to  be

$$H_0\left(p,p^\prime\right) = p\;\delta\left(p
-p^\prime\right)\;\;\;\;\;V\left(p,p^\prime\right) = g\;{{p + p^\prime +
a}\over{p + p^\prime + b}}\;.\eqno (6.1)$$

\vspace{.1in}
\noindent
The interaction term becomes constant in the limit of large $p$ or
$p^\prime$, and the unrenormalized eigenvalue is linear in $\Lambda$.

Applying the Rayleigh-Ritz procedure, we can solve for the bare
interaction term $V_\Lambda\left(p,p^\prime\right)$ in terms of
$V\left(p,p^\prime\right)$.  We will use (4.8) with a lattice in momentum
space as our linearly  independent set of $n$ vectors

$$\chi_i\left(p\right) = \left\{ \begin{array}{ll}
1 & {{i - 1}\over n}\Lambda < p < {i\over n} \Lambda\\
0 &{\rm otherwise}
\end{array}\right.\eqno (6.2)$$

\vspace{.1in}
\noindent
and $F\left(p,p^\prime\right) = \delta\left(p - p^\prime\right)/\left(\mu -
p\right)$.  Thus, the relevant matrices, $A$, $E$, and $C$, are:

\begin{eqnarray*}
A_{i,j} &=& {\Lambda\over n}{{\delta_{i,j}}\over{p_i - \mu}} -
\left({\Lambda\over\ n}\right)^2 {{V\left(p_i,p_j\right)}\over{\left(p_i -
\mu\right)\left(p_j - \mu\right)}} \\
E_{i,j} &=& {\Lambda\over n} \delta_{i,j}\\
C_{i,j} &=&\left({\Lambda\over\ n}\right)^2
V\left(p_i,p_j\right)\left[{1\over{p_i - \mu}} + {1\over{p_j - \mu}}\right]
\end{eqnarray*}

\vspace{-.6in}
$$\eqno (6.3a,b,c)$$

\vspace{.1in}
\noindent
where $p_i = {{i - 1/2}\over n} \Lambda$.  Typical numerical results for $V$
and $V_\Lambda$ are shown in Figures 1 and 2 for $n=100$, $\Lambda=40$,
$\mu=-8.5$, $a=9$, $b=3$, and $g=-2.85$.  Note that $V_\Lambda
\left(p,p^\prime\right)$ has a slightly different functional form in
the regions where only one momentum is large and is almost zero in the
region where both momenta are large.

Using $V_\Lambda$, we can solve Schr{\"o}dingers equation and find the
renormalized eigenvalues and eigenvectors.  If we plot the negative
eigenvalue and eigenvector as a function of $\Lambda$ as shown in Figures 3
and 4, we can see that the system is properly renormalized.\\

\begin{large}
\begin{bf}
\noindent
7.  Local Counterterms
\end{bf}
\end{large}

\vspace{.1in}
An operator that is an arbitrary function of momenta in the momentum-space
basis is generally nonlocal in the coordinate basis.  In light-front
physics, proper renormalization requires that one abandon local operators, at
least for the longitudinal coordinates.  However, when our renormalization
procedure is applied to systems without such infrared divergences, nonlocal
counterterms may still be produced.  This is unacceptable.  One possible
solution to this problem is to choose only functions that have the desired
locality when  choosing a basis of linearly independent operators in the
Rayleigh-Ritz procedure.  Such a basis does not span the space of all
operators,
but the Rayleigh-Ritz procedure will, if possible, choose the coefficients so
that the divergences are still canceled.  While not conceptually pleasing, this
will properly renormalize a system using, by construction, counterterms that
are
sufficiently local.

Let us see how this works in the context of model B.  Since the interaction
term becomes constant in the limit of large $p$, one can renormalize model
B using only a local counterterm.  For simplicity, we will apply the
Rayleigh-Ritz procedure to the counterterm equation for $C_\Lambda$ (2.10)
using
$\left\{O_1\right\}$ as our set of linearly independent operators where $O_1$
is
defined to be $1$ in the momentum space basis.  In this case, we want to find
the stationary point of the functional

$$J\left[ K\right] = {\rm Tr}\; K\left(F + F V F\right)K + {\rm Tr}\; K
\left( F V F V + V F V F\right)\eqno (7.1)$$

\vspace{.1in}
\noindent
where $K = x_1 O_1$; this will give us the best possible approximation to
$C_\Lambda$ from the subspace spanned by $\left\{O_1\right\}$.  Defining

$$B_{1,1} = {\rm Tr}\; O_1 \left(F + F V F\right) O_1 \;\;\;\;\; \tilde{c}_1
= -{\rm Tr}\; O_1\left(F V F V +V F V F\right)\;,\eqno(7.2a,b)$$

\vspace{.1in}
\noindent
the stationary point of (7.1) is given by

$$2 B_{1,1}\ x_1 = \tilde{c}_1\;.\eqno (7.3)$$

\vspace{.1in}
\noindent
In the momentum space basis, we will use $F\left(p,p^\prime\right) =
\delta\left(p - p^\prime\right)/\left(\mu - p\right)$ and,

\begin{eqnarray*}
B_{1,1} & = &-\Lambda \ln {\Lambda\over\mu} + \Lambda \int_\mu^\Lambda
dp\;dp^\prime\; {{V\left(p,p^\prime\right)}\over{p\;p^\prime}}\\
\tilde{c}_1 & = & -2 \int_\mu^\Lambda
dp\;dp^\prime\;dq\;{{V\left(p,p^\prime\right)
V\left(p^\prime,q\right)}\over{p\;p^\prime}}\;.
\end{eqnarray*}

\vspace{-.6in}
$$\eqno (7.4a,b)$$

\vspace{.1in}
\noindent
To leading order in $\Lambda$,

\begin{eqnarray*}
B_{1,1}& = &g \Lambda \left(\ln{\Lambda\over\left|\mu\right|}\right)^2 -
\Lambda\ln {\Lambda\over\left|\mu\right|} - {{\Lambda g
\left(a-b\right)}\over{b+2\mu}}\left\{{{\pi^2}\over 6}+
2\ {\rm Li}_2\left({{b+\mu}\over \mu}\right)\right\}\\
\tilde{c}_1 & = & -2 g^2
\Lambda \left(\ln{\Lambda\over\left|\mu\right|}\right)^2 + {{\Lambda g^2
\left(a-b\right)}\over{b+2\mu}}\left\{{{\pi^2}\over 6}+ 2\ {\rm
Li}_2\left({{b+\mu}\over \mu}\right)\right\}
\end{eqnarray*}
\vspace{-.6in}
$$\eqno (7.5a,b)$$

\vspace{.1in}
\noindent
 where ${\rm Li}_2\left(x\right)$ is the dilnarithm function, and to
leading order in $\ln{\Lambda\over\left|\mu\right|}$,

$$x_1 = -g\left(\mu\right) +  {g\left(\mu\right)\over{1 -
g\left(\mu\right)\ln{\Lambda\over\left|\mu\right|}}}\;.\eqno (7.6)$$

\vspace{.1in}
\noindent
in agreement with references [6] and [7].  Numerically calculating the
eigenvalue
as a function of $\Lambda$ shows that this choice of counterterm does correctly
renormalize the system as shown in Figures 5 and 6 where $n=100$, $\Lambda=60$,
$\mu=5$, $a=8$, $b=2.2$, and $g=-0.65$.

For a more realistic example, one would start with $V$ written as a linear
combination of local operators and use the Rayleigh-Ritz procedure to find
$V_\Lambda$ in terms of those same operators.  We have not yet established
general conditions for the validity of this procedure.\\

\begin{large}
\begin{bf}
\noindent
Conclusion
\end{bf}
\end{large}

\vspace{.1in}
We have introduced an approach to renormalization that is nonperturbative,
general, and successfully handles the UV divergences found in light-front
Tamm-Dancoff field theories.  In this approach counterterms that remove all
cutoff dependence are obtained from the solution of the counterterm equation.
We
have shown that the Rayleigh-Ritz procedure offers a practical way of solving
the counterterm equation, and we have illustrated how the renormalization
procedure works for some simple examples.

In general, the Hamiltonian that one uses in field theory is not the one that
is obtained by canonical quantization. In principle one must include all
operators allowed by power counting.  We have shown that the Rayleigh-Ritz
method can be used to find the combination of allowed operators that provides
the best solution of the counterterm equation. A simple example of this
procedure is given.

Although we have only examined ultraviolet divergences, the discussion
that led to the counterterm equation should also be valid for some of the
infrared divergences found in light-front quantized field theory.  In
light-front coordinates, the Hamiltonian for a free particle with mass $m$,
longitudinal momentum $p^+$, and transverse momenta ${\bf p}_\perp$ is
\cite{perry,yukawa}

$$H_f = {{m^2 + {\bf p}_\perp^2}\over{2 p^+}}\;.$$

\vspace{.1in}
\noindent
Thus, if we associate the $\cal H$ region with {\it small\/} values of $p^+$,
the renormalizability condition can be fulfilled since $H_f\gg E$ for
sufficiently small $p^+$.  Future work will examine the issue of infrared
divergences in detail.

Clearly, there is a lot of work left to be done before we can successfully
describe relativistic bound states using the light-front Tamm-Dancoff
approach.  However, we hope that the approach to renormalization outlined in
this paper will provide a useful framework for addressing the remaining
problems.\\

\begin{large}
\begin{bf}
\noindent
Acknowledgments
\end{bf}
\end{large}

\vspace{.1in}
We would like to thank Judith Gardiner, Ohio State University Department of
Computer Science, for help with numerical issues.  Also, we would like to thank
Robert Perry, Ken Wilson, and Stan G{\l}azek for useful discussion and R.
Jackiw
for pointing out his work on the subject.  We would like to thank Stan
G{\l}azek
and Ken Wilson for a copy of their manuscript on some closely related work.
This work was supported in part by the U. S. Department of Energy.\\

\newpage
\begin{large}
\begin{bf}
\noindent
Figure Captions
\end{bf}
\end{large}

\vspace{.1in}
\begin{description}
\item Figure 1.  Plot of the renormalized interaction term $-V
\left(p,p^\prime\right)$ as a function of $p$ and $p^\prime$ for model B.
\item Figure 2.  Plot of the bare interaction term $-V_\Lambda
\left(p,p^\prime\right)$ as a function of $p$ and $p^\prime$.
\item Figure 3.  Eigenvalue as a function of $\Lambda$ along with a fit to
$x+y/\Lambda$ and the asymptotic value $x$.  The fit parameters are $x=-6.55$
and $y=5.34$. \item Figure 4.  Eigenfunction $p\ \phi\left(p\right)$ as a
function of $p$ for various $\Lambda$.  The wavefunctions are not normalized.
\item Figure 5.  Eigenvalue as a function of $\Lambda$ along with a fit to
$x+y/\Lambda$ and the asymptotic value $x$.  The fit parameters are $x=-6.84$
and $y=30.52$. \item Figure 6.  Plot of the bare interaction term
$-V_\Lambda\left(p,p^\prime\right)$ as a function of $p$ and $p^\prime$.
\end{description}


\begin{thebibliography}{99}
\bibitem{general}More generally, $J\left[ M, K\right] = {\rm Tr}\;M K
\left(F + F V F\right)K - {\rm Tr}\; K\left(F V M + M V F\right)$ is stationary
at $K=V_\Lambda$.  The choice $M=F$ is useful in some cases.
\bibitem{solution}The matrix equation is solved by finding a nonsingular
matrix $U$ such that $U^\dagger A U = {\rm diag}\left(\lambda_1, \lambda_2,
\ldots\right)$ is diagonal and $U^\dagger E U$ is the identity.  The
solution is $X = U {\tilde X} U^\dagger$ where ${\tilde X}_{i, j} =
\left(U^\dagger  C U\right)_{i, j}/\left(\lambda_i + \lambda_j\right)$.
\end{thebibliography}
\end{document}